\documentclass[11pt,a4paper]{article}
\pdfoutput=1
\usepackage{jcappub,amsmath,amsfonts,amssymb,graphicx,subcaption,lineno,afterpage}

\setcitestyle{square}

\begin{document}
%\setpagewiselinenumbers
%\linenumbers
\arxivnumber{1707.01563}

\title{Exploring Potential Signatures of QGP in UHECR Ground Profiles}

\author[a]{Danielle LaHurd}

\author[a]{and Corbin E. Covault}
\affiliation[a]{Case Western Reserve University  \\
10900 Euclid Ave, Cleveland, OH 44106}
\newcommand{\snn}{\mbox{$\sqrt{s_{NN}}$}}
\emailAdd{dvl@case.edu}
\emailAdd{corbin.covault@case.edu}

\abstract{In this work we explore the possibility that the formation
of Quark Gluon Plasma (QGP) during the first interactions of Ultra
High Energy Cosmic Rays (UHECRs) may result in observable signatures
in ground profile and shower particle composition that could
conceivably be detectable by an air shower array experiment such as
the Pierre Auger Observatory.  Knowledge of whether QGP formation
affects the properties of UHECR development will further the
understanding of both UHECR behavior and high energy hadronic
interaction behavior.  We find that for the vast majority of showers
signals of QGP do not manifest themselves in ways that are observable,
but on rare occasion, such as within deeply penetrating showers,
observable signals can be seen.  Results show potential for QGP
detection at 100 PeV initial energy at an initial interaction event
height of 12 km through a $\mu^{\pm}$ excess at 10 GeV between 100 -
300 m from the shower core favoring QGP forming events. In contrast,
higher initial interaction heights of 24 and 36 km at 100 PeV initial
energy show no significant potential for QGP detection.  Unfortunately
at present, the 12 km observable signals cannot be seen with current
detectors, such as the Pierre Auger Observatory\cite{PAOwp}; however,
there may be potential for detection in future experiments.}

\maketitle
\flushbottom

\section{Introduction}

After a UHECR penetrates the atmosphere, it has increasing chances for
interaction as it propagates. Upon the inevitable UHECR-Air collision,
its center-of-mass energy is of equivalent or greater magnitude
(\snn$\sim$100~TeV) to what is currently being run at the LHC.  As a
result, UHECRs and hadronic interactions are fundamentally linked.

It is hypothesized that a new state of matter may exist at very high
energy density, where quarks and gluons become asymptotically free.
This state is called a Quark Gluon Plasma (QGP), in analogy to
electromagnetic behavior of high-temperature collections of charged
particles. A QGP is defined as a local thermal equilibrium, whereby
quarks and gluons are deconfined from hadrons and manifest color
degrees of freedom on nuclear scales rather than nucleon
scales~\cite{STAR05}. The quarks and gluons in a QGP exhibit
fluid-like behavior - "flows" - rather than simpler scatterings that
occur at lower energy density~\cite{Hydro08}.  Signatures of ``flow''
have been reported experimentally from $p + Pb$ collisions at
$\snn=2.76$~TeV collisions at the LHC~\cite{CMS12}~\cite{CMS15}, and
there has been recent evidence of a small QGP formation in $^{3}He+Au$
collisions at RHIC at $\snn=200$~GeV~\cite{HeFlow}.

Since there is recently increasingly strong evidence of QGP formation
at collider energies used at RHIC and the LHC, and UHECR collision
energies are an order of magnitude higher ($\sim 100$~TeV), it stands
to reason that, although the interacting hadrons may be lighter, it is
plausible that QGP formation may be occurring during the initial UHECR
collision.

We note that the question of potentially observable effects of QGP in
UHECRs has been asked in prior work~\cite{Ridky} using a simplified,
two-parton, QGP model, as well as in work~\cite{stringperc} done using
a String Percolation Model.  However, we believe ours is the first
study using up-to-date hadronic interaction models, including
hydrodynamic flow behavior, since first QGP observation in 2005.
Knowledge of whether QGP formation affects the properties of UHECR
development will further the understanding of both UHECR behavior and
high energy hadronic interaction behavior.

% % % % % % % % % % % % % % % % % % % % % % % % % % % %

\section{Methodology}

Our approach is to use simulation studies of high energy interactions
and air shower propagation in the atmosphere to explore the
feasibility that an observational signature indicating the formation
of QGP may be detected based on measurements of ground particles.  In
order for a compelling QGP signal to be measured, several steps in
sequence need to occur:

\begin{enumerate}
\item QGP must be created as a result of first interactions
  between UHECRs and air molecules at a sufficient rate,
\item A sufficient fraction of each QGP must result in child particles
  with characteristics uniquely indicative of QGP, for example,
  multiplicity (N) and flow,
\item A sufficient fraction such events must generate a particle
  cascade where  some imprint of the QGP signature remains in some
  detectable form in the properties of particles arriving to the
  ground, and finally,
\item The distinguishing signatures of ground particles must be
  measurable experimentally and must occur at a rate high enough to
  be detected against any background due to non--QGP cosmic ray
  showers fluctuating so as to mimic the specified signal.
\end{enumerate}

Here, we examine in some detail the plausibility for the first three
steps indicated above.  We begin by applying hadronic
models to primary interactions which allow for the generation of QGP.
We then select a subset of these interactions where QGP signatures in
child particles can be most clearly discerned, operating under the
assumption that only events that generate clear QGP signatures in the
first interaction have any chance to imprint a measurable signature
on the resultant ground particles.  Only the most promising subset of
interactions are then fed into a full air shower simulations which
propagate particles to the ground.  Since the air shower simulations
are most time-consuming computationally, we inject selected showers
into the atmosphere at discrete depths corresponding to three heights:
12~km, 24~km and 36~km spanning a range shower penetration. Finally we
examine distributions of ground particles for signatures of QGP in the
selected air showers relative to a matched set of non-QGP initiated
showers.

Our central aim is not to assess the absolute detectability of QGP in
the context of realistic distributions of cosmic ray and air shower
properties, but rather to assess if any discernible signal might be
seen in ground particles even under optimistic assumptions where
showers are pre-selected with favorable first interactions taking
place deep in the atmosphere.  We emphasize that here that we have not
completed any study to carefully infer the rate at which QGP
signatures might appear in real air shower ground particles generated
from real UHECRs.  Also here we make no attempt to address the extent
to which non-QGP initiated showers can fluctuate so as to yield a
background of air showers that mimic QGP signatures.  Such studies
will require substantially greater computational resources that have
been applied here.

\subsection{Simulation Models} 
\label{Models}

In order to explore the effects of QGP formation during the initial
atmospheric interaction on shower evolution selection criteria are
used on simulated events in order to select for promising QGP
candidate events. A discussion~\cite{FlowpPb14} was consulted to
determine model strengths in simulating and reproducing flow in
$p+Pb$.  The two high-energy hadronic interaction models used for this
discussion are QGSJETII-04~\cite{QGSJET}\cite{QGSmodel} and
EPOS-LHC~\cite{EPOS08}\cite{EPOS09}\cite{EPOSLHC}, both tuned to the
most recent LHC data. QGSJETII-04 does \emph{not} include hydrodynamic
interactions that may occur within the initial hadronic collision
during a QGP formation.  For this reason, we use QGSJETII-04 as the
``non-signal'' comparison for the purposes of this discussion.
Additionally, due to its slightly faster simulation speed, we also use
QGSJETII-04 for the air shower propagation portion of the simulation.

EPOS-LHC~\cite{EPOSLHC} is based on the Parton Gribov-Regge
Theory~\cite{PartonGribov} and includes a parametrized version of
hydrodynamic modeling, replicating QGP effects. A version of EPOS,
EPOS 3.x~\cite{EPOS3}, exists with full 3D+1 viscous hydrodynamic
simulation. However, on consultation with T.~Pierog~\cite{TanguyEmail},
it was determined the compute time required for simulating initial
events in EPOS 3 would be too extreme for the purposes of this study,
with times estimated on the order of a month per each initial event
and no guarantee that the resulting simulated event would have all the
features desired for study. Since the full hydrodynamic simulation of
EPOS 3 is outside our currently available computing resources, the
parametrized hydrodynamics of EPOS-LHC are instead used for simulating
the hydrodynamic, or `QGP signal', initial events.

We separate the modeling of the initial event from the atmospheric
propagation simulation occurring subsequent from the initial
interaction.  Additionally, we use the same model for said atmospheric
propagation for both `signal' and `non-signal' events to limit the
search for differences between the `signal' and `non-signal' to
differences within the initial interaction rather than dealing with
additional atmospheric propagation differences between models.

Specifically, the initial hadronic interactions for both models,
EPOS-LHC and QGSJETII-04, are \emph{not} simulated in the cosmic ray
propagation simulation package, CORSIKA~\cite{CORSIKA}, but are
instead generated in a separate program module called CRMC (Cosmic Ray
Monte Carlo)~\cite{CRMC}, allowing for the simulation of large numbers
of initial events. Selection cuts are made on the CRMC simulated
events before continuing simulation in CORSIKA.

\subsubsection{QGP Event Selection}
\label{selection}
Certain selection cuts on initial collision events have been made in
order to ensure a decent sample size of potential QGP-positive initial
interactions.  All initial interaction events were simulated in 50
event batches using EPOS-LHC and CRMC v1.4 and v1.5.5.  The
simulations were of vertical Neon (Ne) primaries at 100 PeV (0.1 EeV)
hitting a Carbon (C) target at rest ($\snn= \sim3$~TeV comparable to
LHC energies).  Neon was chosen due to two factors: simulation time
and nucleon density. Iron (Fe) and Silicon (Si) primaries were also
tested. While an iron primary would be the ideal for testing behaviors
and observables due to increased density, the computing time required
with our current available resources made it untenable. A neon primary
sits in the ``sweet spot'' of compute time for available resources,
around 2 - 4 days per simulation, with a reasonable probability for
observing QGP behavior in initial interactions while still being a
potential, if rare, cosmic ray primary candidate. Protons, while
having the fastest compute time, only demonstrate QGP behaviors in the
most extreme high multiplicity circumstances. These high multiplicity
(N $>$ 1800 particles) initial events occur in $<1\%$ of simulated
proton events. The initial energy of 100 PeV was also chosen due to
compute time restrictions.

We are aware that collisions of 100 PeV cosmic rays may probe just the
onset of quark deconfinement, and air shower signals for the formation
of a quark gluon plasma could still appear at much higher energies. It
is important to probe the entire energy parameter space from the onset
of deconfinement and higher energies to determine what effects
deconfinement has on shower evolution and development.

\begin{figure}[!htbp]
		\centering
	\begin{subfigure}[t]{.3\textwidth}

		\includegraphics[width=0.8\textwidth, angle=-90]{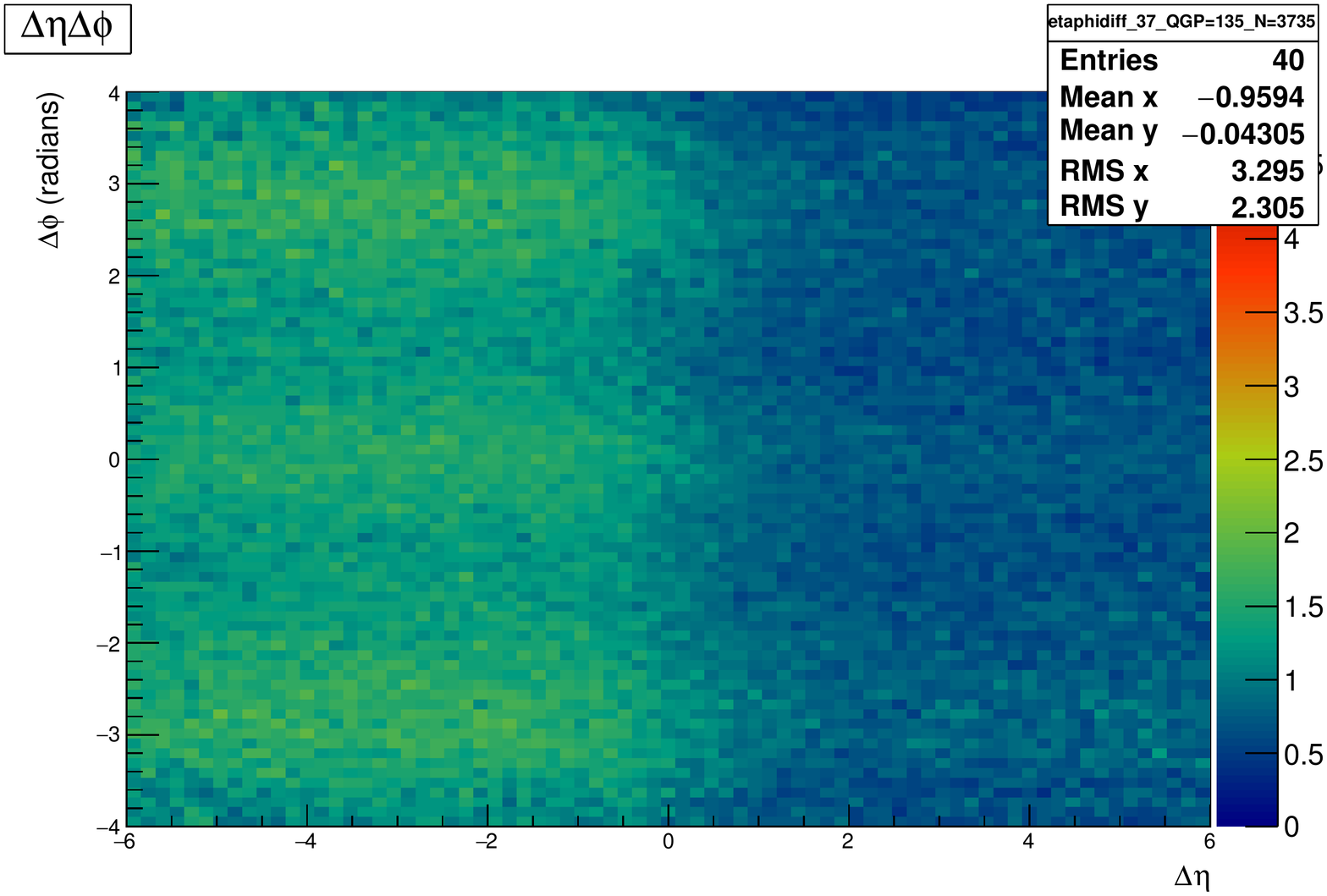}
		\caption{EPOS-LHC initial condition with QGP effects present.  } %Note the collective effect ``ridges'' appearing near $\Delta\phi$ = 0 and $\Delta\phi=\pm\pi$.}
		\label{fig:EPOS}
	\end{subfigure} %
    \hfil %
    %\medskip
	\begin{subfigure}[t]{.3\textwidth}
		\includegraphics[width=0.8\textwidth, angle=-90]{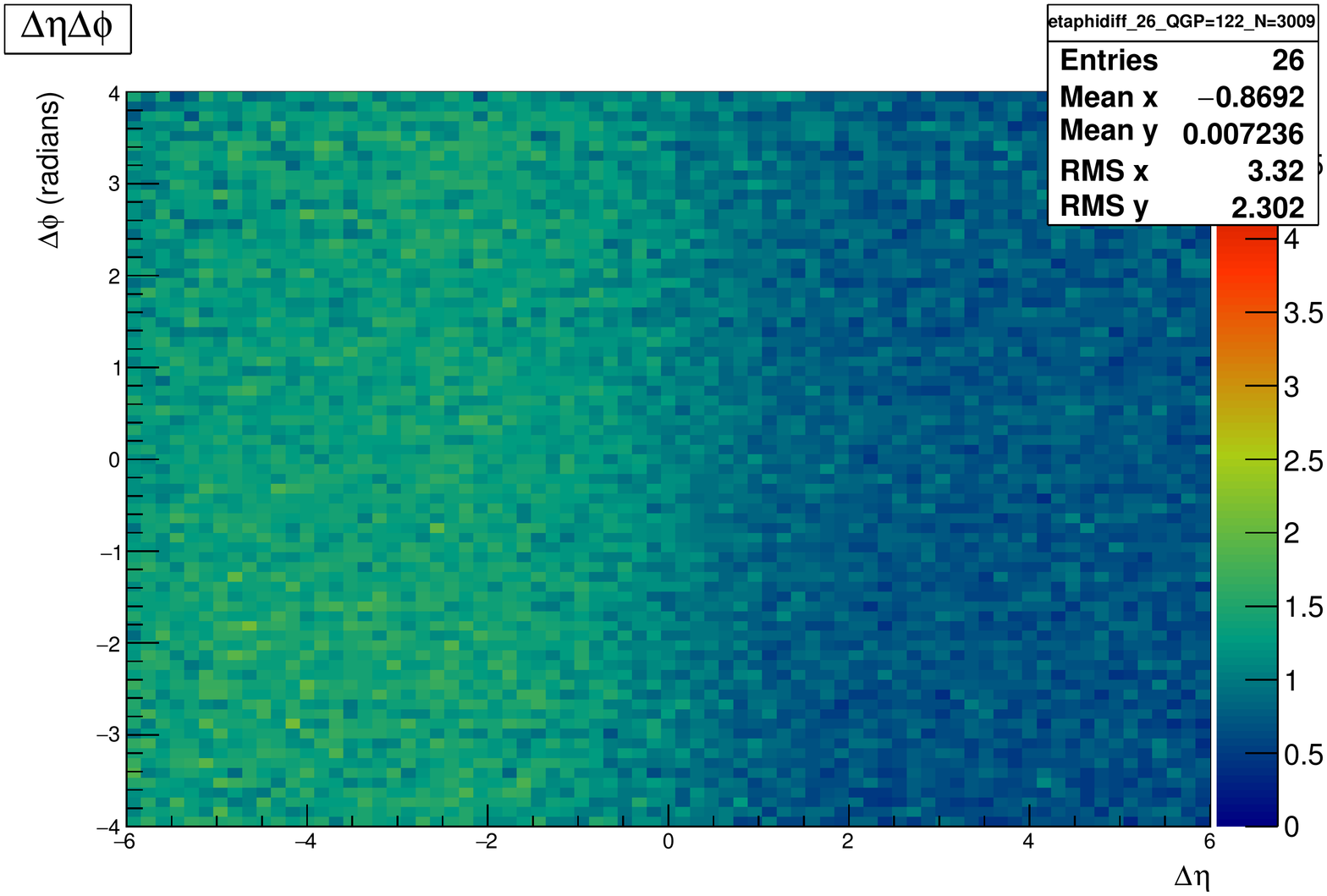}
		\caption{EPOS-LHC initial condition without full QGP presence. }% There is some collection at negative $\Delta\eta$ but no ``ridges'' present.}
		\label{fig:EPOS-NO}
	\end{subfigure} %
    \hfil% %
	\begin{subfigure}[t]{.3\textwidth}
		\includegraphics[width=0.8\textwidth, angle=-90]{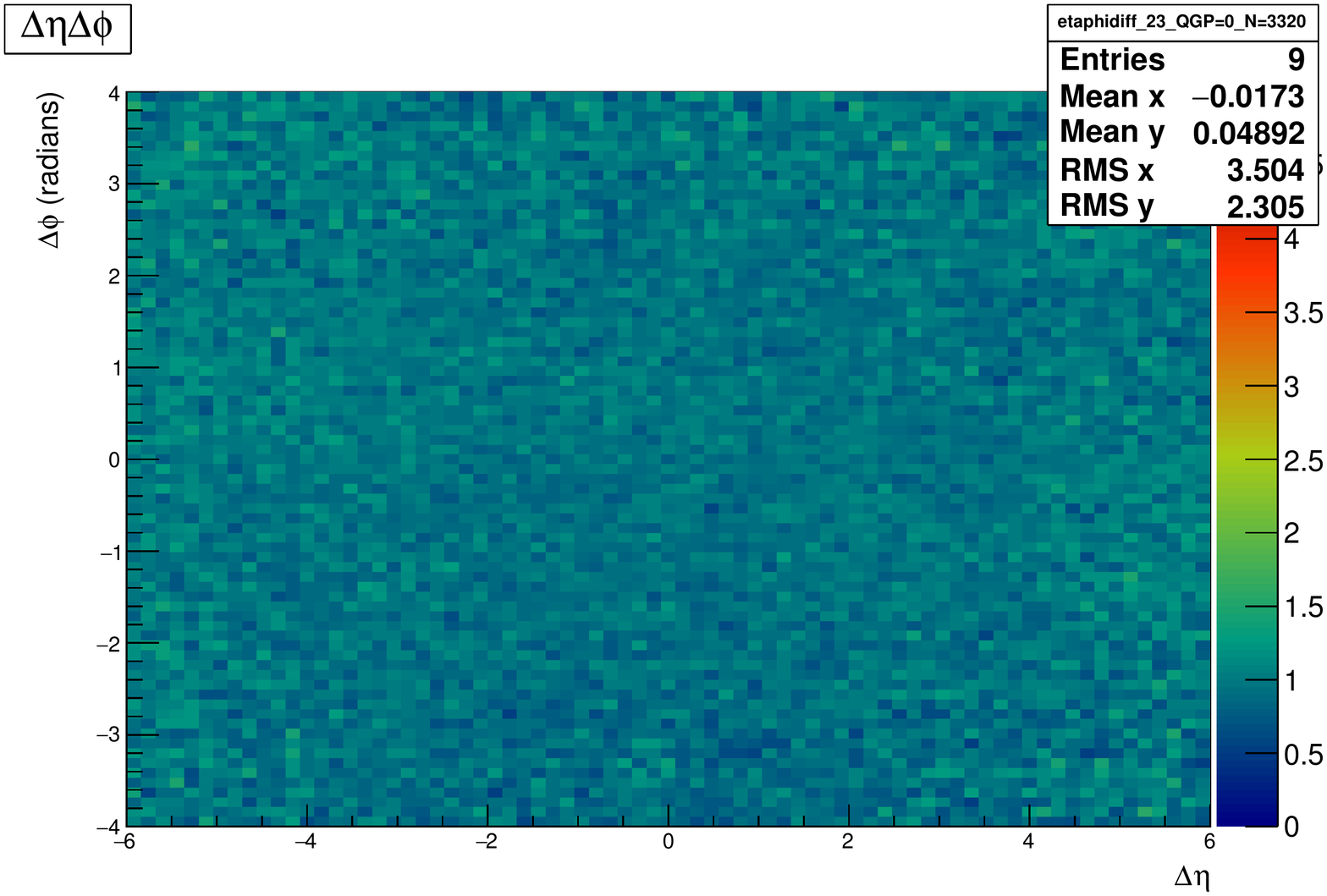}
		\caption{QGSJET initial condition.  Note no collective behavior at any $\Delta\eta$ or $\Delta\phi$.}
		\label{fig:QGS}
	\end{subfigure}
    %\begin{minipage}[t]{0.3\textwidth}

	\caption{Simulated particle distributions for pairs of
          particles produced in under different hadronic interactions
          models as a function of both psuedo-rapidity difference
          $\Delta\eta$ and azimuth angle difference $\Delta\phi$.  The
          color code shows the number of particle pairs found for a
          given $\Delta\eta$ and $\Delta\phi$.  All events have $\sim
          3000$ particles.}  % \end{minipage}
		\label{fig:diffplots}        
\end{figure}

%  QGSJETII-04 simulated events are considered non-signal events
%  and do not show collective behavior, as is expected.
Before any full $Ne+C$ simulation was done, a small series of test
showers were run to compare outputs between EPOS-LHC and QGSJETII-04.
Figure~\ref{fig:diffplots} shows the azimuthal angular and pseudorapidity
difference of pairs of particles produced in hadronic interactions
simulated with and without collective effects. The color code shows
the number of particle pairs found for a given difference in azimuthal
angle and pseudorapidity.  EPOS-LHC events showed signs of collective
behavior (\ref{fig:EPOS} and \ref{fig:EPOS-NO}), while, as expected,
the QGSJETII-04 events did not show the same collective behavior
(figure \ref{fig:QGS}). Similarly, when QGP effects are turned off in
EPOS, the results are similar to the QGSJETII-04
output~\cite{Miller07}.

All QGP candidate events are selected to have more than 2000 particles
at initial freeze-out. This requirement selects for high multiplicity
events and reduces false positives in $\Delta\eta\Delta\phi$
graphing. Events with lower than 2000 freeze-out particles generally
lack sufficient statistics to discern any flow-like effects.  Events
with greater than 4000 particles at freeze-out were also cut, due to
both the rarity of these events ($<0.004\%$) and difficulties in
observing possible flow behavior without additional pseudo-rapidity
($\eta$)\footnote{$\eta = -\ln \left( \tan\frac{\theta}{2}
  \right)=\frac{1}{2}
  \ln\frac{|\textbf{p}|+p_{z}}{|\textbf{p}|-p_{z}}$} cuts via the
large number of particles obscuring the flow effects. In principle,
additional $\eta$ or $p_{T}$ cuts could be made to view flow effects
within the high multiplicity candidates; however, for this analysis,
these cuts were not made.

The QGSJETII-04 ``non-signal'' initial events are also given the same
multiplicity cut, limiting the events to those with
$2000<N<4000$. QGSJETII-04 initial events tend to have higher initial
multiplicity due to the lack of hydrodynamical effects and QGP density
induced suppression. We limit the multiplicity of QGSJETII-04 events
to ensure similar initial conditions to those of the QGP `signal'
events.

%impact parameter
The impact parameter, or separation of the centers of the two nuclei
when they collide, is provided by the HepMC~\cite{HepMC} output of a
CRMC simulation. Only QGP candidate events that have a low impact
parameter, i.e. high nucleus overlap, are selected for this study as
low impact parameter correlates strongly with high multiplicity. The
impact parameter chosen was $b<5$ fm, which corresponds to roughly
0\%-15\% centrality (85\% - 100\% nucleus overlap).

%V2 cosine fits
In order to obtain events with high ``flow'' corresponding to a
possible QGP signature, the events are graphed on a $\Delta\phi$ vs. N
graph, then fitted with a Fourier cosine function describing flow:
\[
f(\Delta\phi) = 1+\sum_{n=1}^{3} 2c_{n}cos(n\Delta\phi)
\]

It is the $v_{2}$~($n$=2) term that is of most interest, as it is a
sign of strong collective behavior that is influenced positively by
the presence of a QGP.  Fits were used to select for strong QGP
candidate events. If the $v_{2}$ coefficient, corresponding to
elliptic flow, was below 0.02, the event was rejected, as lower
$v_{2}$ values correspond to weaker flow effects (low initial
anisotropy). This fitting selection process ignores fit errors and is
only used as a framework for selecting potential QGP events.

The pass rate with all cuts is about 1.4\%. Table \ref{tab:cuteff}
shows a summary of the selection cuts applied and the overall
efficiency.

\begin{table} %Cut efficiency table
        \begin{center}
            \begin{tabular}{ | l | l |l|}
            \hline
            \textbf{Cut Applied} &\textbf{Number of Events} &\textbf{Percent Remaining}\\ \hline
            Total events simulated & 3550& 100\% \\ \hline
            Impact Parameter $<$ 5 fm & 1918& 54\%\\ \hline   
            Multiplicity cut of N $\geq$ 2000 & 538 & 15\%\\ \hline
            Multiplicity cut of N $<$ 4000 & 524 &14.7\%\\ \hline
            Cosine fit $v_{2}>0.02$ and visible `ridges' & 51 & 1.4\% \\ \hline          
            \end{tabular}
            \caption{The number of EPOS-LHC events simulated for potential QGP events and the efficiency of applied selection cuts.}\label{tab:cuteff}        
        \end{center} 
\end{table}  

\section{CORSIKA Simulation}
\label{Simulation}
%Thinning parameters
While EPOS-LHC and QGSJETII-04 are used for the high energy hadronic
interaction simulations, it is the program CORSIKA~\cite{CORSIKA} that
takes these models and uses them to model the propagation of particles
through the atmosphere. The version used for the hadronic interaction
comparison simulations at the time of the discussion is CORSIKA
v74004. CORSKIA transitions to a lower energy interaction model, FLUKA
(v2011.2)~\cite{fluka03}\cite{fluka05}, for particle simulations when
the energy of individual particles falls below 100 GeV (configured at
time of simulation).

Thinning in CORSIKA is the process by which computing time is
shortened by only following one particle from a cascade below a
certain energy rather than every particle individually. Thinning does
not preserve flavor counts or baryon numbers~\cite{TanguyEmail},
therefore care must be taken when deciding the thinning range so as
not to lose valuable data.  All CORSIKA showers simulated at 100 PeV
were thinned at the 10 GeV level. There is assumed to be no
discernible signal in the particles below 10 GeV as additional
atmospheric interactions and secondary showers will have clouded the
signal.

\subsection{``Head'' and ``Body''}
For the ``head'' or initial interaction, EPOS-LHC and QGSJETII-04
initial events are chosen, as described above in section
\ref{selection}, and formatted into a CORSIKA readable format.

For the air shower simulation, or ``body'', the QGSJETII-04 model was
chosen for atmospheric simulation of both initial interaction models.
This is to allow for only the initial interaction type to influence
the developments within the air shower and allow for more direct
comparisons. QGSJETII-04 was chosen over EPOS-LHC for the ``body''
simulation as it requires slightly less time for simulation.  All
showers were simulated using the same version of CORSIKA(v74004) and
FLUKA to reduce systematic errors.

\subsubsection{Converting to CORSIKA Format}

\begin{figure}[!htbp]
\centering
\includegraphics[width=0.9\linewidth]{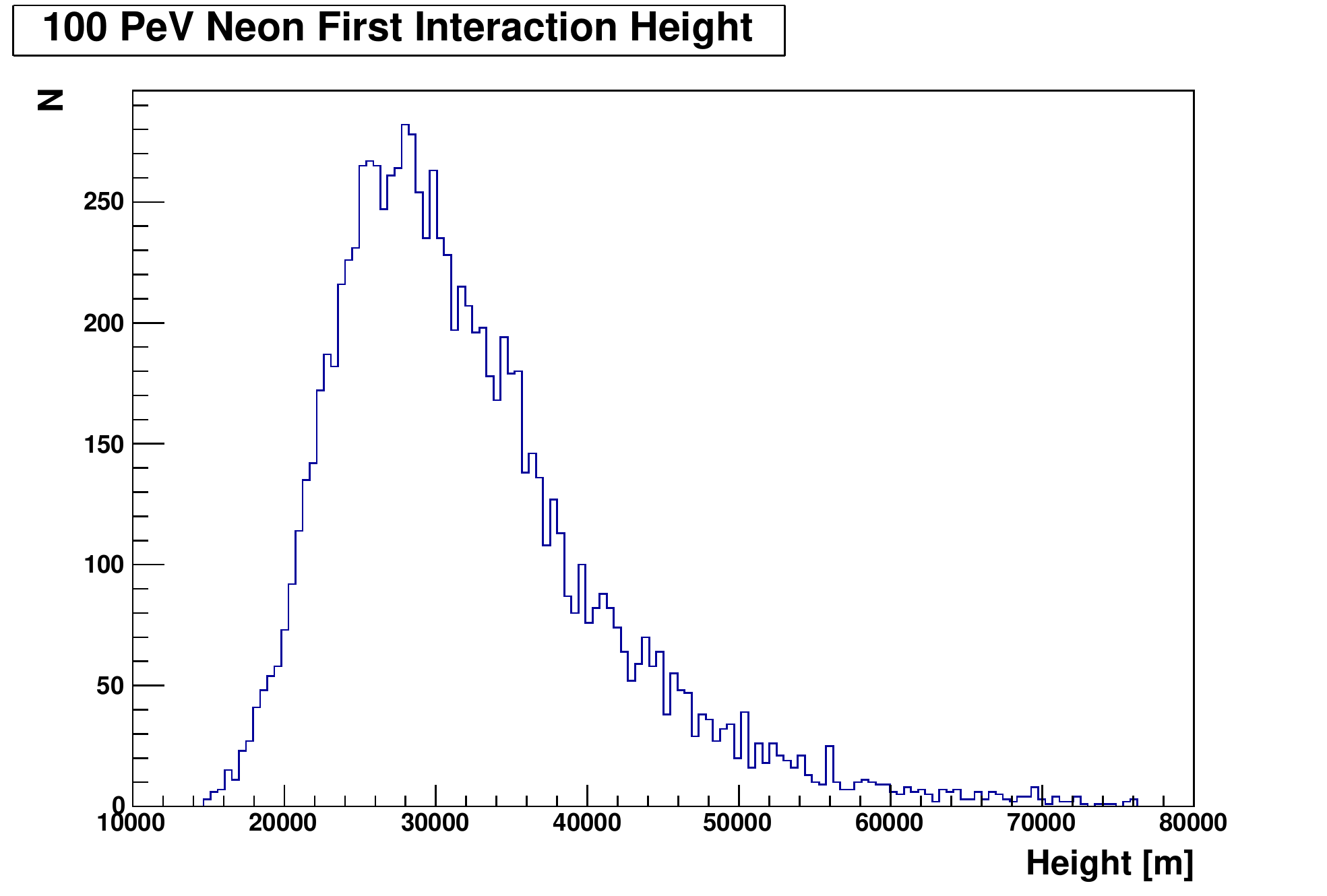}
\caption{Graph initial interaction height of 100~PeV neon primary showers through 10000 CORSIKA simulations.}
\label{fig:heights}
\end{figure}

%heights
As the shower is not being generated internally by CORSIKA, an initial
height from the detector plane must be provided for the air shower
simulation to begin. Many, very thinned, test showers have been
simulated in CORSIKA using proton, carbon, and iron primaries in order
to determine the typical initial collision height. For this
discussion, the initial collision heights have been chosen at discrete
values of 12, 24, and 36 kilometers. For detectability, 12 km
potentially shows the most signal originating from the initial
collision due to less atmosphere attenuation since first
interaction. However, as seen in figure \ref{fig:heights} with neon
primary interaction heights, it would be extremely uncommon to see an
air shower, especially one of somewhat heavier composition such as the
neon primaries used, originating at such an extreme atmospheric
penetration depth. Interactions set to 24 km represent a value close
to the average expected initial height of a shower with a composition
near carbon or neon mass, and 36 km represents a high
starting-elevation shower.  A total of 51 showers have been simulated
for each model (102 total showers) for this discussion, with 17
showers generated at each of the three initial heights for both
models.

\subsection{Analysis}

The CORSIKA output files are processed, using C++ scripts, into a ROOT
format file consisting of compact particle data such as position,
species, generation, time of impact, and momentum.

During analysis we make an energy cut of all particles below 10
GeV. Most particles below 10 GeV result from secondary showers and
decays which can wash out potential signals from initial interactions
that we wish to examine. Additionally, making the cut at 10 GeV allows
us to eliminate particle weights introduced by CORSIKA though
simulation thinning. The removal of the weighted particles can be
beneficial as the thinning and weighting of particles, while vastly
speeding up simulation time, only preserves energy and not baryon or
lepton numbers~\cite{TanguyEmail}. By not preserving baryon and lepton
numbers amongst the weighted particles, the particle species
abundances may be altered in a non-physical way and may hide potential
differences between the evolution of the two initial interaction
types.

The 17 simulations for each model and height are averaged and
normalized before comparison to obtain a single distribution for
initial interaction model comparison purposes.

One of the ways we compare the two interaction models is through
taking the difference between the normalized distributions.  This
provides a way to look for QGP dependent excesses or deficiencies,
after atmospheric evolution.  In order to determine the significance
of any differences between the QGP and non-QGP results, we divide the
differences by the propagated errors of the difference.
\[ \mathcal{R} \equiv
\frac{\langle N_{QGP}\rangle- \langle N_{noQGP}\rangle}{\sigma}
\] 
Here $\cal R$ is the ``residual excess'', a measure of the number of
standard deviations the difference deviates from a null-hypothesis
scenario where both QGP and non-QGP initiated air showers produce the
same results, and $\sigma$ is the RMS uncertainty in the difference
between QGP and non-QGP values.

The statistical significance of any one difference is, of course,
diluted by the fact that we are searching for a range of different
potential signatures. Although the ``trials factor'' is difficult to
estimate, \emph{a posteriori}, we conservatively expect our study to
correspond to several hundreds of independent searches.  For the
purposes of initial analysis, we therefore assign an $\cal R$ value
greater than 3$\sigma$ as a cause for interest, and assign an $\cal R$
value of greater than 5$\sigma$ to represent a likely compelling
physical difference between the models.  We plot distributions of
$\cal R$ for various different measurements. A distribution centered
near zero, signifies that both initial interaction models evolve
similarly, or at least do not display significant differences after
the initial interaction. If the distribution deviates far from zero,
or there are bins residing significantly outside the central
collection, there may be an excess in favor of either the QGP
(positive) or non-QGP (negative) events.

Additionally, we preform a histogram comparison between 1D variables
of QGP and non-QGP modeled events using the $\chi^{2}$ test of
homogeneity~\cite{chi2}.  We examine the resulting normalized
residuals, i.e. the residuals divided by their standard deviation,
from the comparison. If QGP formation has no detectable influence on
shower development the residuals should remain distributed at or near
zero.  Should any residuals bin deviate strongly from zero, it
indicates a potential region where the presence of QGP during the
initial collision has influenced shower development.  The magnitude of
the residuals' deviation from zero scales with, but is not equivalent
to, the significance of said deviation.

% % % % % % % % % % % % % % % % % % % % % % % % % % % % % % % %

\section{Results}

For the presented results we will primarily focus on the 12 km
simulated events, as both the 24 and 36 km events show little
differences between QGP (EPOS-LHC) and non-QGP (QGSJETII-04) models
with the number of events simulated.

\subsection{Initial height: 12 km}
\label{12km}

Due to the depth of atmospheric penetration required, a 12 km initial
(first interaction) height is an extremely unlikely condition for a
Ne, or heavier, primary at 100 PeV.  However, examining results at
this height provides a informative tool for determining whether
viewing QGP effects from the initial interaction is feasible after
traversing the atmosphere, or if any potential signal will be
eliminated by the numerous interaction lengths traveled. The 12 km
sample represents an overly optimist sample which is the most
favorable for the detectability of QGP signature observables.  Should
no observables be present on the ground after the comparatively short
distance of 12 km there will be little hope of finding detectable
observables at the higher, more realistic, initial interaction heights
of 24 and 36 km.

\begin{figure}[!htbp]

\begin{subfigure}[t]{.48\textwidth}
    \centering
     \includegraphics[width=0.96\linewidth,page=1]{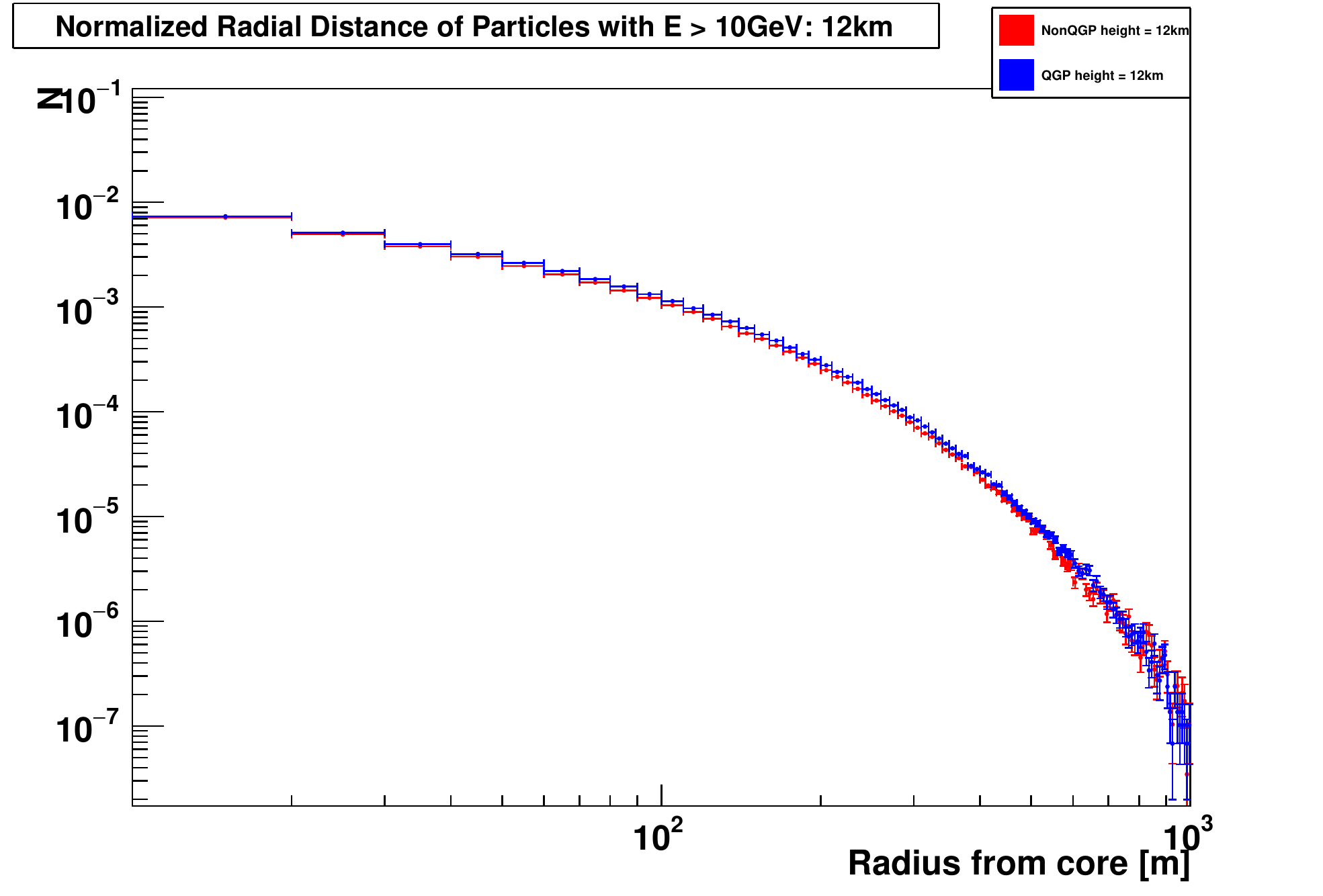}
    \caption{Normalized radial distribution: QGP excess near 100 m.}
    \label{fig:12kmR}
\end{subfigure} %
\hfil%
\begin{subfigure}[t]{.48\textwidth}
    \centering
    \includegraphics[width=0.96\linewidth,page=2]{1D_bigfont_trim.pdf}
    \caption{Normalized residuals: QGP excess up to 300 m.}
    \label{fig:12kmRres}
\end{subfigure} %
\hfil%
\begin{subfigure}[t]{.48\textwidth}
    \centering
     \includegraphics[width=0.96\linewidth,page=3]{1D_bigfont_trim.pdf}
    \caption{Significance values: Centered on the positive axis; multiple bins are above 5$\sigma$.}
    \label{fig:12kmRsig}
\end{subfigure}
\begin{minipage}[b]{.48\textwidth}

\caption{Radial distribution comparison of all particles with energy exceeding 10 GeV between QGP and non-QGP events with an initial interaction height of 12 km.}\end{minipage}
\end{figure}

We examine the radial distribution of particles from the 12 km events
for any differences between the distributions of the QGP and non-QGP
`headed' showers.  Upon viewing this radial distribution for all
tracked particles (figure \ref{fig:12kmR}) we note that QGP (blue)
events demonstrate an excess of particles at a 100 m distance from the
core.  This is difficult to see in the initial log-log comparison,
however, until we examine the residuals (figure \ref{fig:12kmRres})
comparing the QGP results to the non-QGP results. In the residuals,
the QGP excess (positive) is clearly visible from 50 m up to about 300
m.  This difference appears to be significant (figure
\ref{fig:12kmRsig}) as there are multiple significance bins exceeding
5$\sigma$. Additionally, the significance distribution trends towards
positive significance, indicating a overall QGP excess.

\begin{figure}[!htbp]
\begin{subfigure}[t]{.48\textwidth}
    \centering
    \includegraphics[width=0.96\linewidth,page=1]{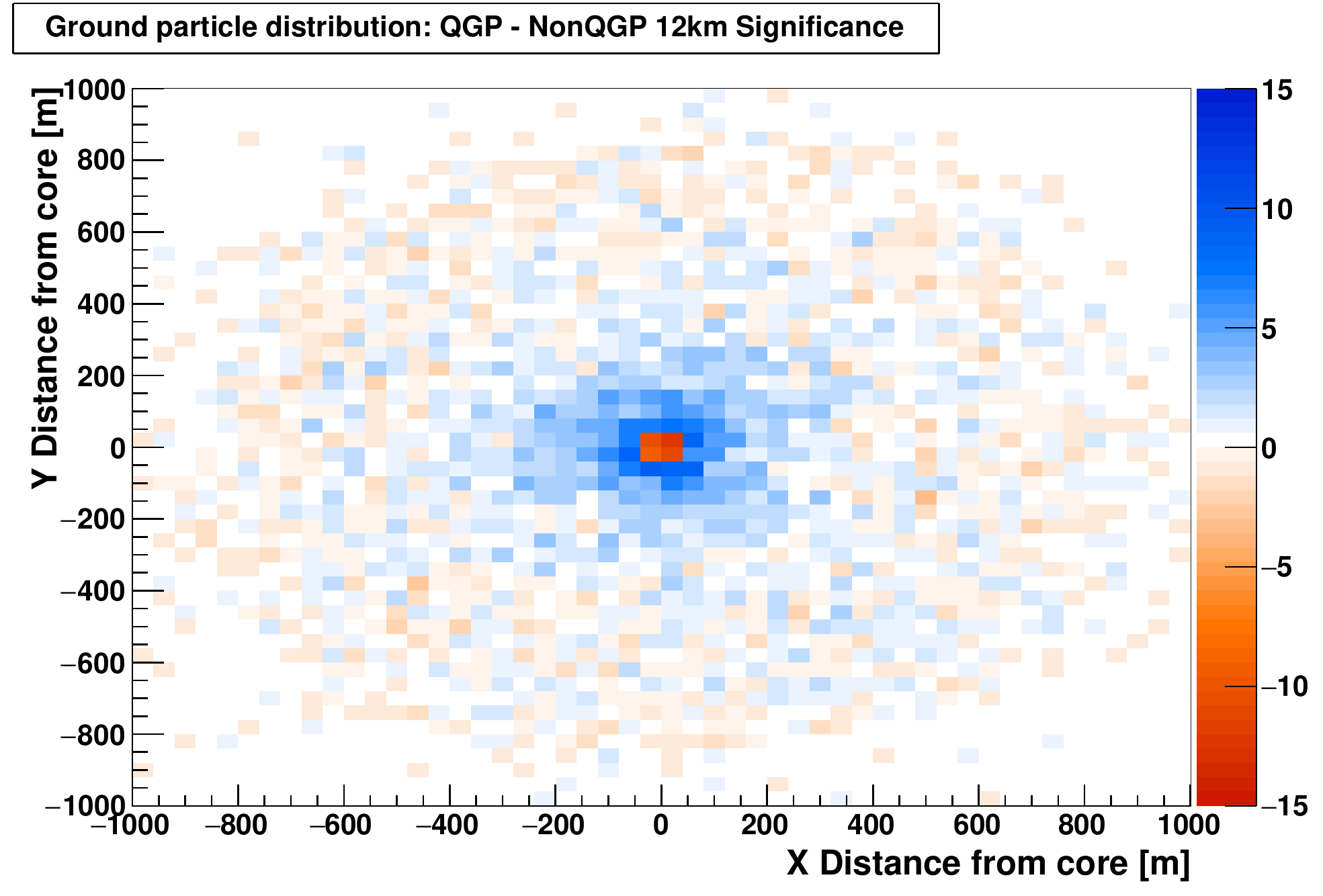}
    \caption{Significance($\sigma$) of the difference of normalized counts by $(x,y)$. }
    \label{fig:12kmXYsig}
\end{subfigure} %
\hfil%
\begin{subfigure}[t]{.48\textwidth}
    \centering
    \includegraphics[width=0.96\linewidth,page=2]{2D_bigfont_trim.pdf}
    \caption{Distribution of significance values of the difference of normalized counts.}
    \label{fig:12kmXYsigdist}
\end{subfigure}
\caption{Particle distribution $(x,y)$ with energy exceeding 10 GeV and an initial interaction height of 12 km. Significance favoring QGP is positive(blue) and non-QGP is negative(red)}
\end{figure}

Examining a top-down view of the particle distribution, this
QGP-favored excess between 50 m and 300 m remains significant when
distributed across the azimuthal bins (figure
\ref{fig:12kmXYsig}). The significance values from this distribution
(figure \ref{fig:12kmXYsigdist}) also exceed 5$\sigma$.  There may be
an event-to-event azimuthal clustering; however, this was not studied
in this discussion.

\begin{figure}[!htbp]
\begin{subfigure}[t]{.48\textwidth}
    \centering
     \includegraphics[width=0.96\linewidth,page=1]{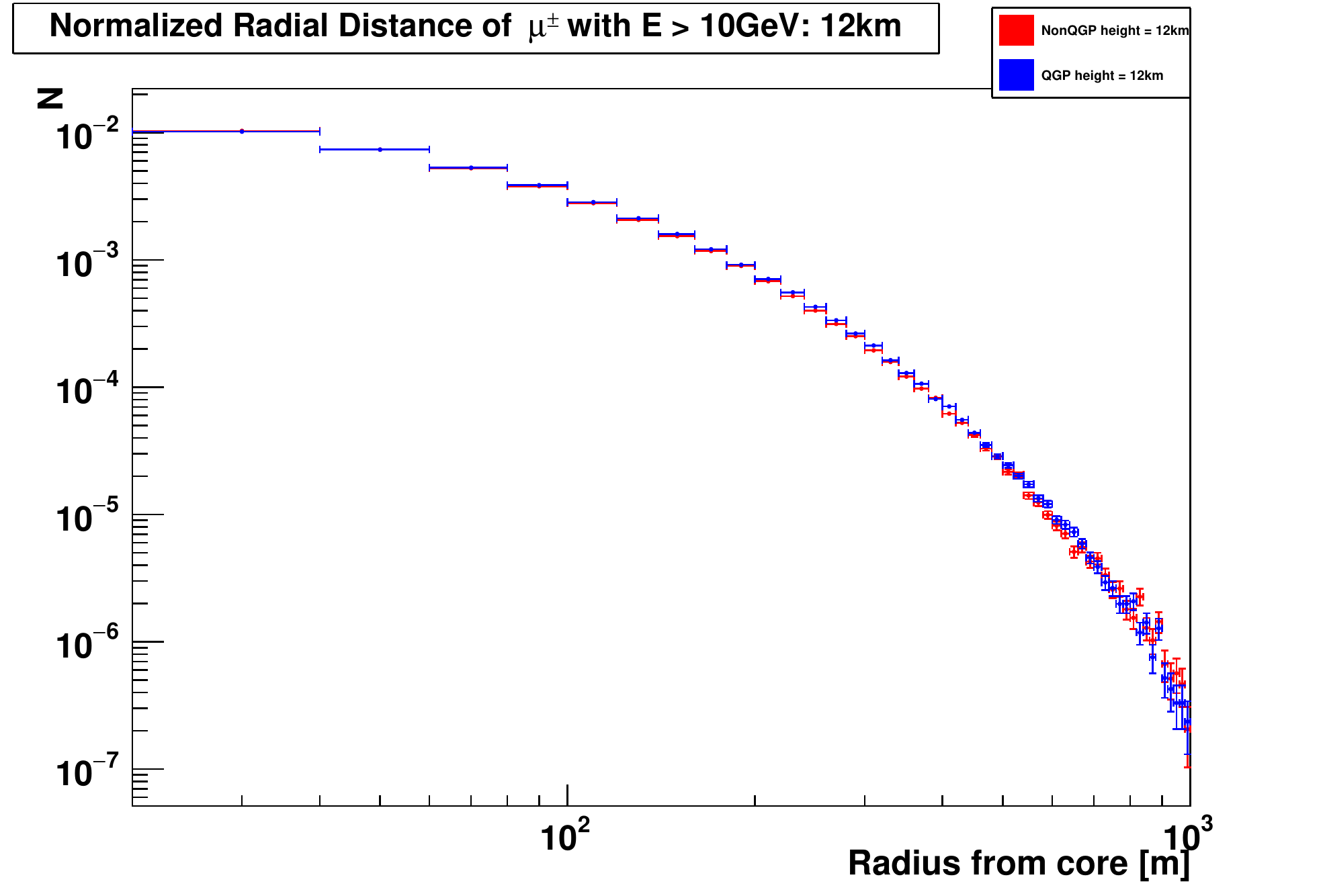}
    \caption{Normalized radial distribution: QGP events are in blue and non-QGP events are in red.}
    \label{fig:12kmRmu}
\end{subfigure} %
\hfil% %
\begin{subfigure}[t]{.48\textwidth}
    \centering
     \includegraphics[width=0.96\linewidth,page=2]{1Dmu_trim.pdf}
    \caption{Normalized residuals: QGP-favored excess between 50 and 300~m.}
    \label{fig:12kmRmures}
\end{subfigure} %
\hfil% %
\begin{subfigure}[t]{.48\textwidth}
    \centering
      \includegraphics[width=0.96\linewidth,page=3]{1Dmu_trim.pdf}
    \caption{Significance values}
    \label{fig:12kmRmusig}
\end{subfigure} %
\hfil% %
\begin{minipage}[b]{.48\textwidth}
\caption{Radial distribution comparison of $\mu^{\pm}$ with energy exceeding 10 GeV between QGP and non-QGP events with an initial interaction height of 12 km.}\end{minipage}
\end{figure}

Attempting to untangle which particles are contributing to the
QGP-favored excess, we examine particle distributions separately based
on particle species. The muon profile (figure \ref{fig:12kmRmu})
displays an excess in favor of QGP. This excess is once again
corroborated by the residuals (figure \ref{fig:12kmRmures}) exhibiting
an excess of $\mu^{\pm}$ in favor of QGP between 50 and 300 m.  This
excess is not as statistically significant (figure
\ref{fig:12kmRmusig}) as that seen in the full particle distribution
but does contain a number of bins with 5$\sigma$ significance.
On examination, the remaining individual particle species radial
distributions show little to no differences between the two
models. Therefore, we conclude that the muons dominate the apparent
QGP-favored particle excess seen a 50 - 300 m from the core.

\begin{figure}[!htbp]
\begin{subfigure}[t]{.48\textwidth}
    \centering
    \includegraphics[width=0.96\linewidth]{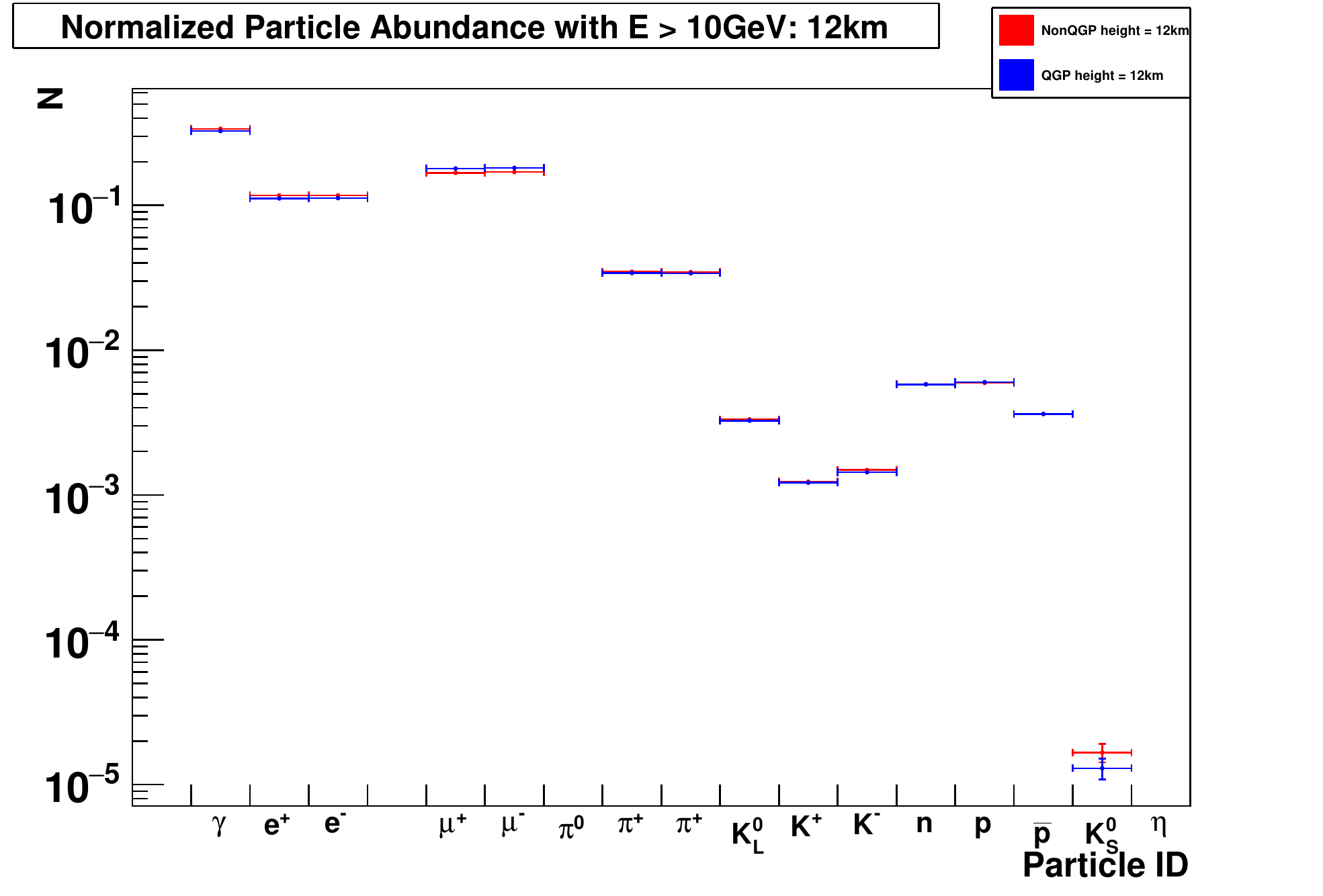}
    \caption{Normalized counts of particle species. QGP events are in blue and non-QGP events are in red.}
    \label{fig:12kmID}
\end{subfigure} %
\hfil% %
\begin{subfigure}[t]{.48\textwidth}
    \centering
     \includegraphics[width=0.96\linewidth,page=4]{1D_bigfont_trim.pdf}
    \caption{Normalized residuals comparing particle species.}
    \label{fig:12kmIDres}
\end{subfigure}
\caption{Comparison of particle species counts and significance with energy exceeding 10~GeV and an initial interaction height of 12~km.}
\end{figure}

As there appears to be a potential signal in muons for differentiating
QGP and non-QGP showers, we examine particle species abundances within
the entire ground profile for other possible difference between the
models.  A normalized count of particle species (figures
\ref{fig:12kmID} and \ref{fig:12kmIDres}) shows a significant
difference between the two models for muons and EM. There is little
difference between QGP and non-QGP for the other tracked particle
species. The EM ($\gamma,e^{\pm}$) excess strongly favors non-QGP
events and $\mu^{\pm}$ excess strongly favors QGP events.

\subsubsection{Initial height: 12 km Discussion}
There appears to be a significant difference, especially in the radial
particle distribution, in the muon and EM output between the QGP and
non-QGP `heads' at 12~km.  For the non-QGP `headed' events, there is a
$e^{\pm}$ excess seen within the core region when compared to QGP
headed events.  However, there is no current detector capable of
measuring the direct shower core output due to the required fine-grain
surface detector spacing of $\sim$100~m$-$300~m.  As such, an
$e^{\pm}$ core excess or deficiency cannot be used as a signature to
determine whether an initial event has formed QGP using previously
collected data from experiments past or existing, such as Auger.

As for muons, there is a significant difference between the initial
interaction model outputs when viewing the area between 50 and 300 m
from the core.  However, this is only apparent when comparing the
discrepancy to the outer regions of the ground profile, where both
models are identical.  While the spacing of currently existing
detectors is not ideal, this signature is potentially detectable if
one uses a tightly clustered detector with $\sim$100 m spacing.  To
search for a QGP signature, one would have to compare the peripheral
of the shower's ground profile with the profile of the 50 to 300 m
region and find an anomalous excess of particles in comparison to
other shower profiles.  This could be made easier with the addition of
scintillation panels to water Cherenkov surface detectors to better
separate muon particle detections from electron detections.

%\afterpage(\clearpage)

\subsection{Initial height: 24 km}
\begin{figure}[!htbp]
\begin{subfigure}[t]{.48\textwidth}
    \includegraphics[width=0.96\linewidth,page=3]{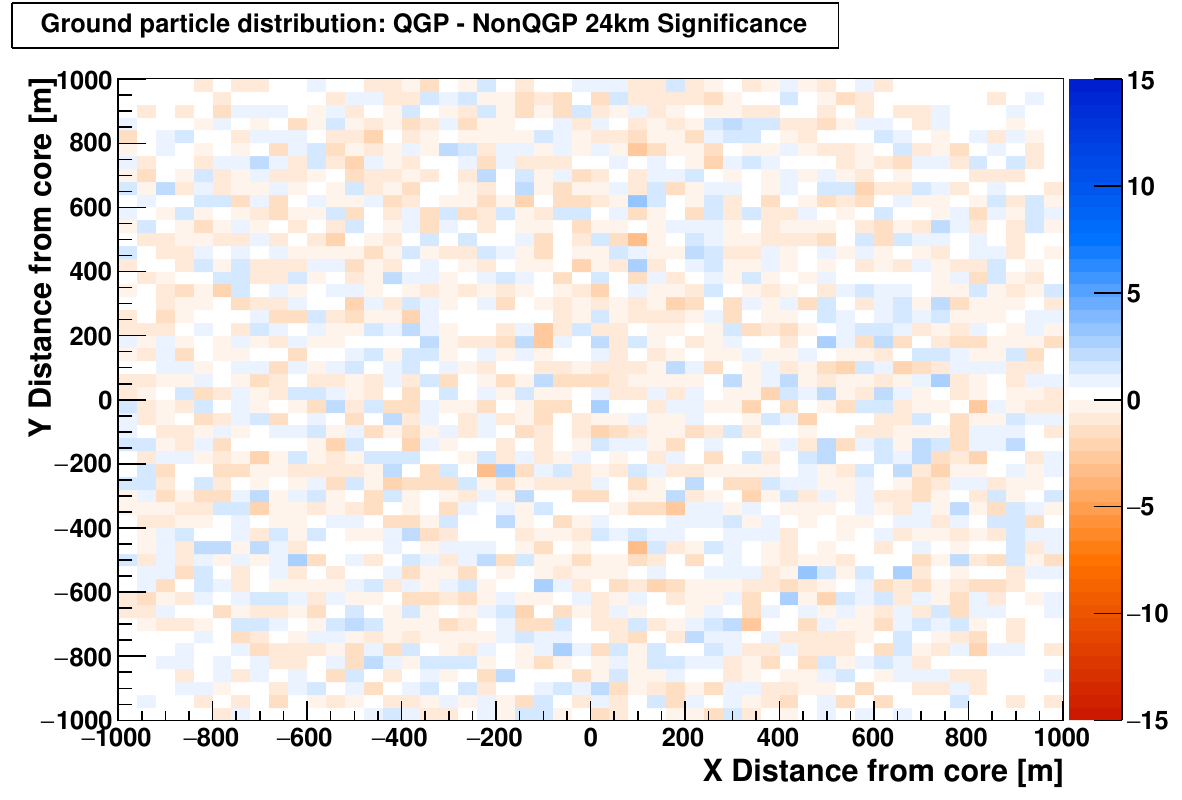}
    \caption{Plot of the significance of the difference in particle count on an $x-y$ plane. There is no favoring of either model present.}
    \label{fig:24kmXYdif}
\end{subfigure} %
\hfil % %
\begin{subfigure}[t]{.48\textwidth}
    \includegraphics[width=0.96\linewidth,page=4]{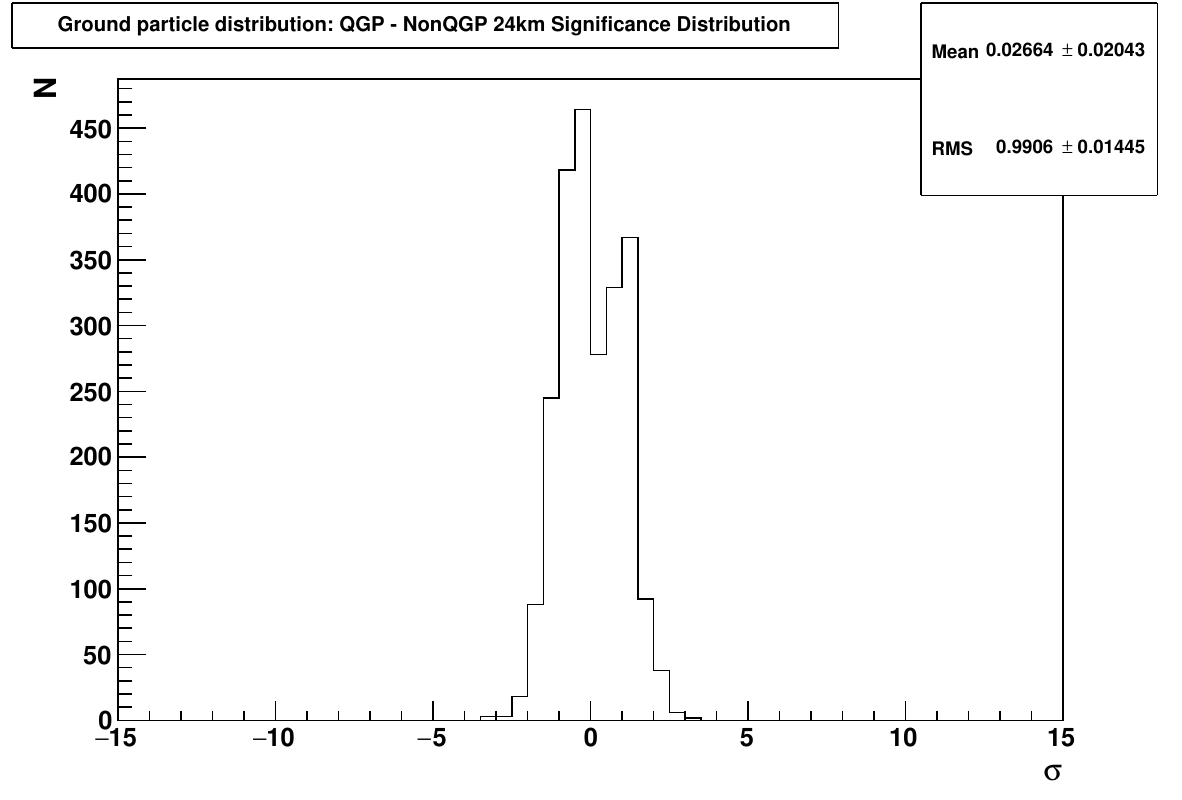}
    \caption{Plot of significance value distribution. The values are centered on zero with no deviations outside 3$\sigma$ favoring one side over the other.}
    \label{fig:24kmXYsig}
\end{subfigure}
\caption{Significance of the difference in distribution of all particles with E $>$ 10~GeV between QGP and non-QGP events with initial height of 24~km. Positive (blue) signifies a distribution in favor of QGP, while negative (red) signifies a distribution in favor of non-QGP.}
\end{figure}

For 24 km initial heights, looking at the top-down view of the
particle distribution (figure \ref{fig:24kmXYdif}) there appears to be
no evidence of excess.  In fact, the $(x,y)$ distribution of all
particles exceeding 10 GeV, with a 24 km initial height, appears to
show no difference between the two models in particle count. This is
confirmed by the near-Gaussian distribution of the significance values
in figure \ref{fig:24kmXYsig}.

%\afterpage(\clearpage)
\subsection{Initial height: 36 km}
\begin{figure}[!htbp]
    \centering
\begin{subfigure}[t]{.48\textwidth}

    \includegraphics[width=0.96\linewidth,page=5]{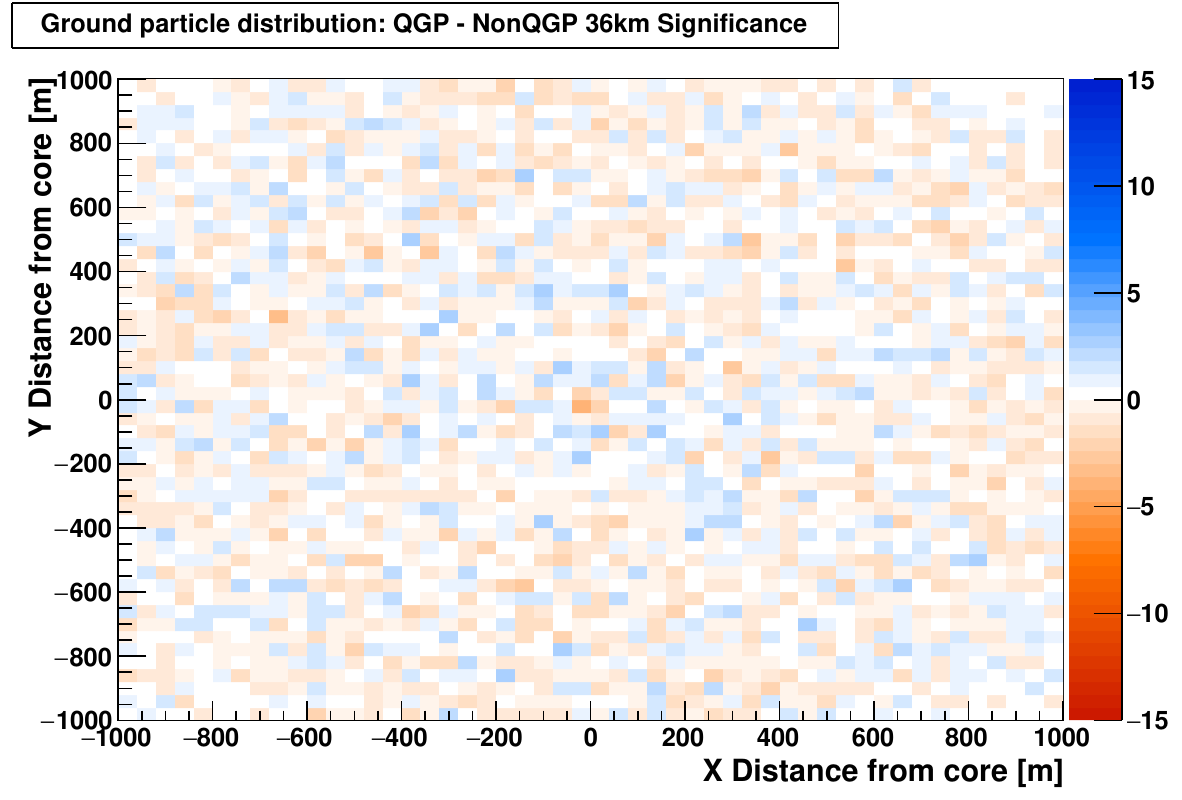}
    \caption{$(x,y)$ plot of the significance ($\sigma$) of the difference in distribution. Note the randomness of the distribution, signifying no favoring of either model.}
    \label{fig:36kmxysig}
\end{subfigure} %
\hfil% %
\begin{subfigure}[t]{.48\textwidth}
    \includegraphics[width=0.96\linewidth,page=6]{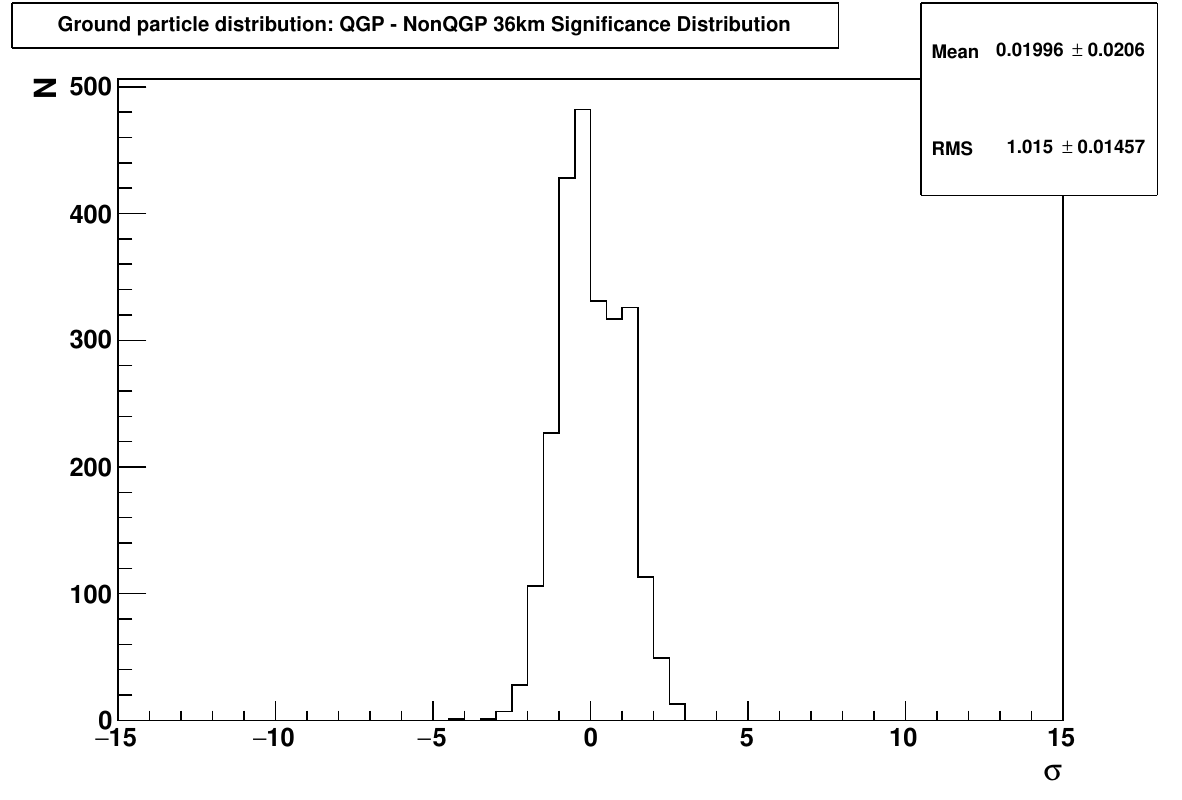}
    \caption{Plot of the significance distribution. There are no outliers and the significance is centered near zero, implying similar distributions.}
    \label{fig:36kmxysiggaus}
\end{subfigure}
\caption{Significance of the difference in distribution of all particles with E $>$ 10 GeV between QGP and non-QGP events with initial height of 36 km. Positive (blue) signifies a distribution in favor of QGP, while negative (red) signifies a distribution in favor of non-QGP.}
\end{figure}

Any differences between QGP and non-QGP events as measured from the
ground particle profile (figures \ref{fig:36kmxysig} and
\ref{fig:36kmxysiggaus}) at a 36 km initial height appear to be
statistically negligible and similar to the 24 km radial distribution
discussed previously. Neither figure demonstrates a favoring of one
model over the other but rather display an extreme similarity between
a QGP initiated shower and a non-QGP initiated shower.  This may be a
result of too much time, and/or too many interaction lengths, since
first interaction eliminating any perceivable differences.

Unfortunately, the lack of statistically significant effects at 36 km
is discouraging for prospects of detecting QGP in real UHECR air
showers as a 36 km initial height is many times more likely to occur
than one at 12 km. This is especially true with the penetration
potential of heavier primaries at 100 PeV, which would be more likely
to have QGP formation during initial collision due to containing more
interacting nucleons.  Ironically, the most favorable conditions for
QGP formation occur with the least likely ability for detection, as
neon and similar weighted primaries are most likely to interact at a
higher elevations.

% % % % % % % % % % % % % % % % % % % % % % % % % % % % % % % % % % % % % % % % % % % % % % % % %

%\afterpage(\clearpage)
\section{Discussion}
The aim of this discussion is to determine if there exist effects on
the ground profile and particle composition of a UHECR air shower due
to the formation of QGP during the initial interaction of the UHECR on
the atmosphere. The knowledge of whether QGP formation affects the
properties of UHECR development would enhance understanding of both
UHECR and high energy hadronic interaction behaviors.

Based on the results, the conclusion we reach is, in all practicality,
detecting whether QGP has formed within a UHECR is extremely
challenging. While there is potential in the 12 km results, more
simulated events are needed to verify the extent to which the features
seen in the 12 km results remain both present and detectable. With 102
total events simulated, and only 34 at each height, a statistical
anomaly within in a single event can alter the perceived results. A
more reasonable number of simulations for review would on the order of
100 simulated events per model at each height, thereby limiting the
effects of anomalies to 10\% or less rather than a 25\% effect with
the current number of events.  Simulating on the order of 1000 events
per height and model would only be feasible with supercomputing
resources, but would limit the effects of any outlier or anomalous
events to $\sim$3\%.

We must also consider the extreme rarity of a QGP formation event
occurring. Via our selection methods, a QGP with potentially
detectable properties, e.g. strong hydrodynamic flow and reasonable
multiplicity, has a $\sim$1.4\% chance of occurring with a neon
primary at 100 PeV.  These chances are lower for lighter primaries due
to fewer interacting nucleons limiting chances for high multiplicity.

Moreover, the probability of a of a primary particle penetrating
deeply enough to have a 12 km initial height is quite small.  Based on
the height distribution presented in figure \ref{fig:heights} for a
neon primary at 100~PeV, a 12 km initial height for particles of
similar mass is a very rare occurrence with only a 3.18\% chance for
neon to penetrate deeper than 20 km and a 0.15\% chance below 16
km. As a reminder, the 12 km height was primarily chosen as a `proof
of concept', whereby a lack of any observables at said height would
prove that QGP formation is undetectable in UHECRs.

Therefore, due to the combination of both the rarity of detectable
events and the lack of detectability at higher elevations, we must
conclude the possibility of observing QGP formation from a UHECR is
very small, below a 0.0021\% chance per shower, at 100 PeV with a neon
or similar mass primary event with an ideal (simulated) detector that
records all ground particles.

Although possible detectability is slim at a less than 0.0021\% chance
per shower, with the upcoming Pierre Auger Observatory
upgrade~\cite{AugerUpgrade} there is potential for detecting
deeply-penetrating, QGP-forming, showers.  The upgrade adds an
additional scintillation panel on top of the currently existing Auger
water Cherenkov surface detectors, allowing for more accurate electron
and muon discrimination, to a resolution of 15\% or better in
determining the number of muons~\cite{AugerUpgrade}.  Additionally,
the upgrade allows for measurement of shower properties nearer the
shower core due to reducing the number of saturated PMTs, from a 1000
m radius to 300 m. The now 300 m radius lies within the potential
signal region produced in the simulated QGP-forming events.

As seen from the 12 km results, the finer spacing and muon
discrimination in the Pierre Auger upgrade may be enough to detect
signs of QGP formation, as the region where signals appear is muonic
at 300 m from the core.  Care must be taken to select for the deepest
penetrating showers which can be inferred either from coincidental
fluorescence measurements or through lateral density profiles.
Confirmation of results may be limited, however, due to the rarity of
the specific conditions required for detectability.

For example: An effective area of the Pierre Auger Observatory Infill
(where the spacing between detectors is reduced from 1.5 km to 750 m)
region of about 50 km$^{2}$ would yield approximately 100,000 events
during a 5-year operating window.  At a 0.0021\% rate, this would
result in about 2 QGP-detectable events based on selection cuts.
Although this is in principle a detectable number of events, the
rarity of QGP means that selection cuts against non-QGP showers must
have very high rejection factors.

\section*{Acknowledgments}
This discussion was made possible by NSF grant PHY-1207523. D.~LaHurd
acknowledges the support of the Timken Fellowship at Case Western
Reserve University. Data analysis was made possible in part by
CRMC\cite{CRMC}.  We acknowledge very helpful discussions with and
guidance from T.~Pierog. We are grateful for editorial and scientific
comments and suggestions on the manuscript from J.~Matthews and other
members of the Pierre Auger Collaboration.
% % % % % % % % % % % % % % % % % % % % % % % % % % % % % % % % % % %
\clearpage

\bibliographystyle{plain}

\bibliography{bibliography_tryagain}
\end{document}